\documentclass[multphys,vecphys]{svmult}

\usepackage{makeidx}         % allows index generation
\usepackage{graphicx}        % standard LaTeX graphics tool
\usepackage{multicol}        % used for the two-column index
\usepackage[bottom]{footmisc}% places footnotes at page bottom
\makeindex             % used for the subject index
\begin{document}

\title*{AO assisted spectroscopy with SINFONI: PSF, background, and
  interpolation}
\titlerunning{AO spectroscopy with SINFONI}

\author{R. Davies\inst{1}}

\institute{Max Planck Institut f\"ur extraterrestrische Physik, 85741,
  Garching, Germany
\texttt{davies@mpe.mpg.de}}
\maketitle

I discuss 3 widely applicable aspects concerning calibration of the
near infrared adaptive optics integral field spectrometer SINFONI:
(1) the accuracy with which one needs to quantify the PSF and how this
might be achieved in practice; 
(2) how it is possible to fine tune the background subtraction to minimise
the residual OH airglow;
and 
(3) how an altered perspective on calibration data might lead
to improvements in interpolation and greater flexibility in
reconstructing datacubes.

\section{A Short Introduction to SINFONI}
\label{davies:sec:intro}

SINFONI \cite{davies:eis03a,davies:bon04} is a versatile instrument
comprising of a 60-element curvature adaptive optics system
\cite{davies:bon03} that feeds a 1--2.5\,$\mu$m integral field
spectrometer \cite{davies:eis03b}.
The camera has 3 pixel scales spanning 0.25$^{\prime\prime}$ to
0.025$^{\prime\prime}$, making it adaptable to both seeing and
diffraction limited and resolutions.
The associated fields of view range from
$8^{\prime\prime}\times8^{\prime\prime}$ to 
$1^{\prime\prime}\times1^{\prime\prime}$.
It can cover the H and K bands together in a single exposure at a
spectral resolution of $R\sim1500$;
or a complete single waveband (J, H, or K) at $R\sim2000$--5000,
depending on the pixel scale.
Since the highest resolution (associated with the smallest pixel
scale) is under-sampled, one has the option of spectrally dithering
and interleaving the 2 exposures.
Image slicers dissect the field of view and re-arrange the slitlets
along a single pseudo-slit.
On the detector, the dispersed data from each slitlet appear exactly
analogous to standard longslit data, except that 
there are 32 such 2D spectra next to each other.
A dedicated data reduction package {\em spred} \cite{davies:abu05}
reconstructs the 3D datacube.
An excellent tool for viewing them is {\em QFitsView} (see
\verb+http://www.mpe.mpg.de/~ott/QFitsView+), which displays the
spectrum in real time as one moves the pointer across the spatial
field. 
With this tool it is also extremely quick and simple to apply a wide
range of processing techniques in real time.

\section{The Adaptive Optics Point Spread Function}
\label{davies:sec:psf}

Misunderstandings about adaptive optics PSF abound:
that it must be known in great detail, and that its temporal and
spatial variability casts doubt on interpretation of the data.
In this section, I attempt to alleviate these concerns by
discussing some ideas about the level of accuracy with which one needs
to know the PSF, and some ways in which this might be achieved.

\subsection{Quantifying the PSF}

There will always be some situations where it is necessary to know the
PSF in great detail,
most obviously in planet searches where one is trying to
detect small faint object close around a bright point source.
In these cases, it is crucial to distinguish between the
object and structure that belongs to the PSF.
On the other hand, many -- perhaps most -- applications do not require
such a detailed level of knowledge.
Particularly for extragalactic science, where the AO correction is
mediocre, a simple combination of 2 analytical functions will often
suffice.
For example, the PSF can be generally well matched by the sum of a
narrow Gaussian, which represents the core of the PSF, and a Moffat
function which can trace the wide wings in the halo.
Greater detail in the PSF is unnecessary
because the accuracy is limited by the model (kinematic or
morphological), which, in contrast to the real intrinsic structure in
galaxies, is usually simple, and often symmetric.

\subsection{(De)-convolution}

\begin{figure}[t]
\centering
\includegraphics[width=2.7cm]{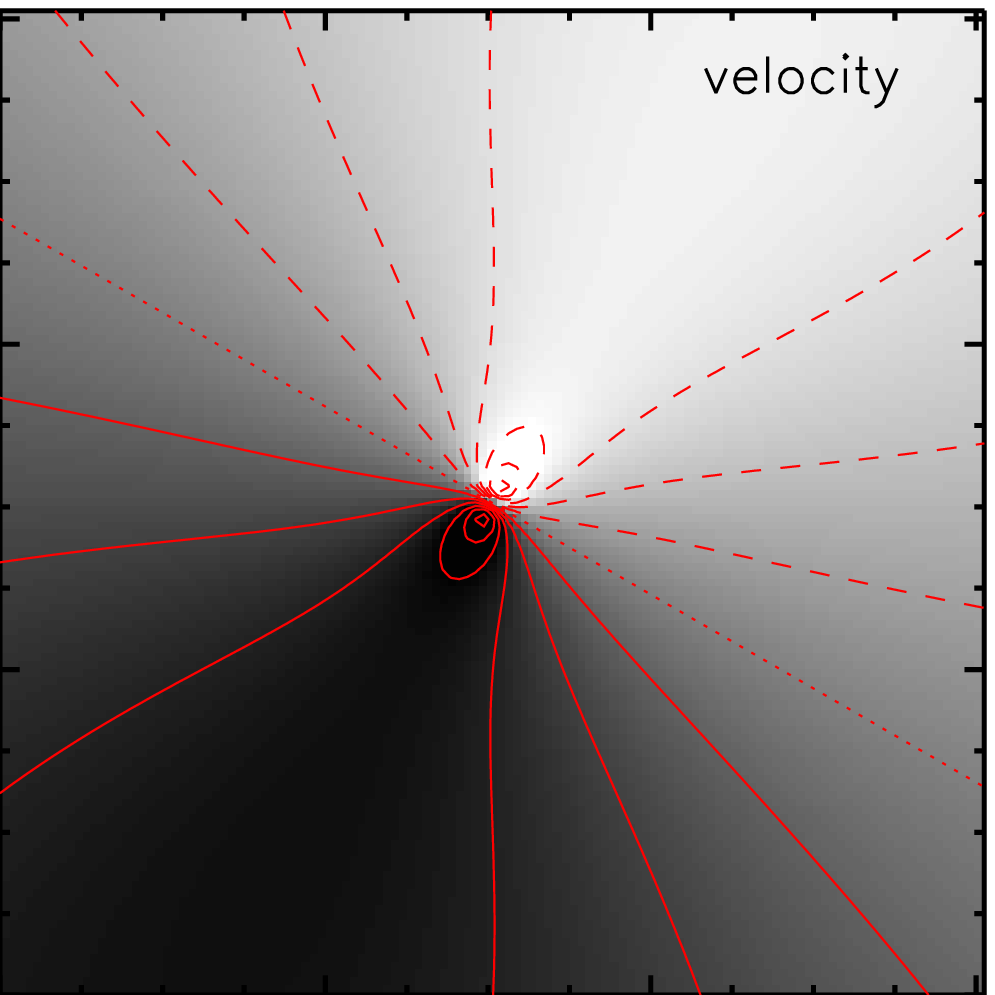}
\includegraphics[width=2.7cm]{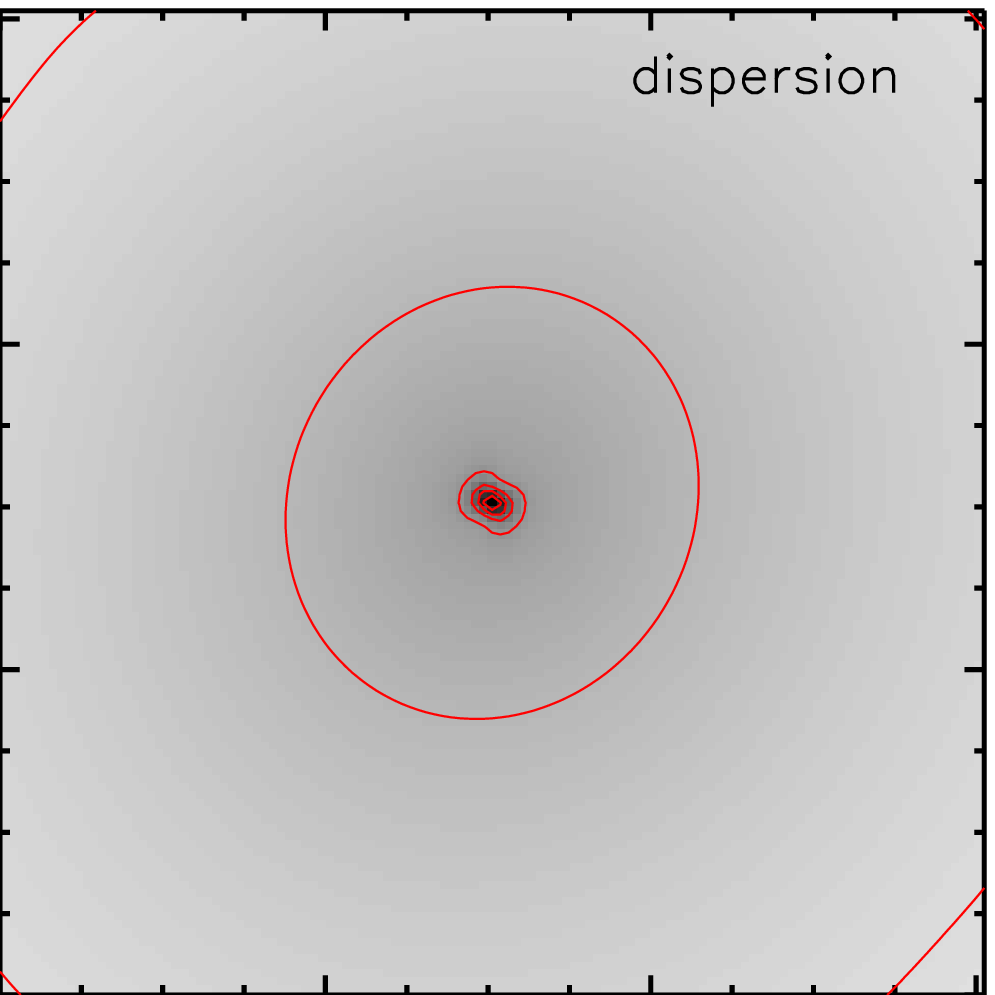}
\hspace{5mm}
\includegraphics[width=2.7cm]{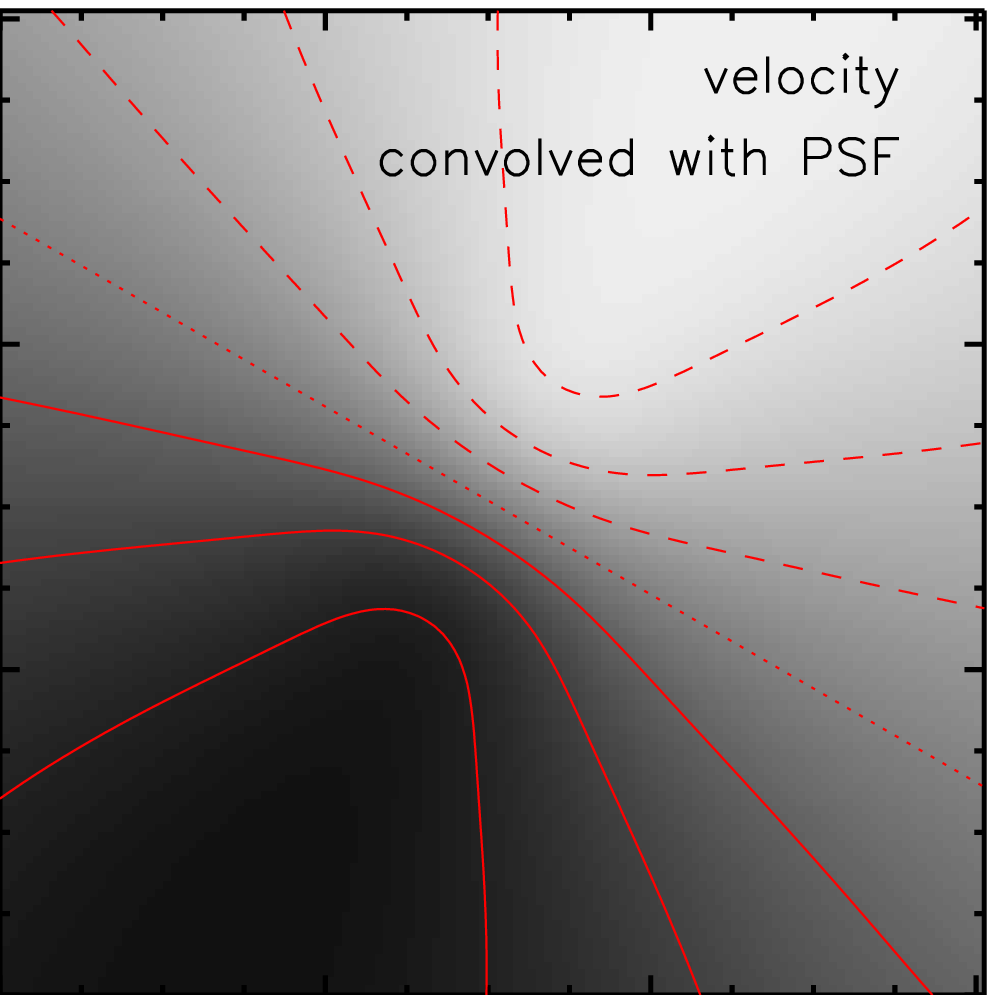}
\includegraphics[width=2.7cm]{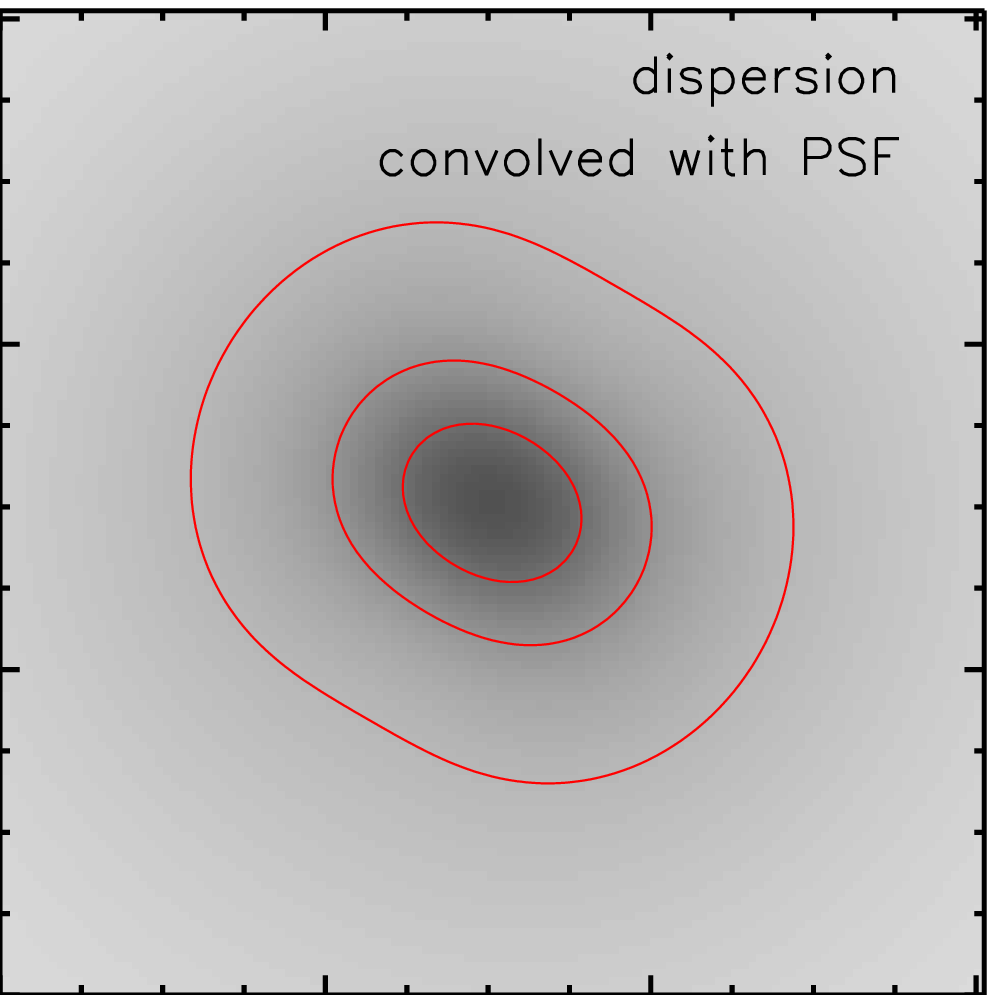}
\caption{Velocity and dispersion fields for matter in a
  self-gravitating disk around a supermassive black hole. In the field
  shown, the integrated mass of the disk is 5 times that of the black
  hole. The effect of PSF-induced smearing on the kinematics is very
  dramatic. Due to cross-talk between these 2 quantities
  (i.e. velocity gradients on scales comparable to the PSF contribute
  to the dispersion), the 
  smeared kinematics cannot be deconvolved. The only option is to
  create a 3D kinematic model (2 spatial and 1 velocity dimension),
  convolve it with the PSF, and then extract the kinematics. By
  iterating one can constrain the model parameters.
}
\label{davies:fig:kin}       % Give a unique label
\end{figure}

Dealing with a PSF that comprises 2 contrasting
components -- a narrow core and broad wings -- is an important
issue.
Clearly one needs to separate the PSF from the instrinsic structure in
the observed data.
But deconvolution is not always the best solution.
It is an inverse problem, and hence mathematically messy, 
tends to amplify noise, and can easily generate unreal artifacts
(e.g. ringing).
And at the end, one still has no convenient expression for the
intrinsic source shape and one still has to deal with a PSF in the
deconvolved data.
It may be narrower, but it is probably less well defined and may
vary with signal-to-noise across the field.
In some cases, such as kinematics
(Fig.~\ref{davies:fig:kin}), deconvolution is simply not an option due
to the cross-talk between the velocity and dispersion.

\begin{figure}[t]
\centering
\includegraphics[width=11cm, clip=]{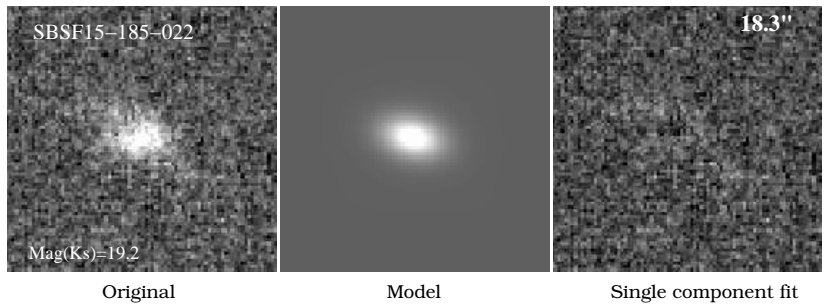}
\caption{
Example of a faint $K_s=19.2$ galaxy observed with adaptive optics
\cite{davies:cre05}.
The galaxy was 18.3'' from the guide star, and so the PSF derived for
it had to take into account isoplanatic effects (see point (iii)
below).
The image of the galaxy (left) is too noisy to deconvolve with the PSF. 
It has therefore been fitted by convolving the PSF with a
parameterised galaxy profile (centre). 
The parameters of the profile were adjusted to minimise the residuals
(right).
This can yield not only the best fitting parameters, but also a good
estimate of their uncertainties.
}
\label{davies:fig:cresci}
\end{figure}

An alternative is to convolve a model of the intrinsic structure with
the PSF, compare the result to the observations, and adjust the model
iteratively.
This is the basis of popular galaxy fitting algorithms, such as galfit
\cite{davies:pen02} which has been used in Fig.~\ref{davies:fig:cresci}.
The method also enables one to make realistic estimates of the
uncertainty in the fitted parameters.
While it cannot be used if there is no way to parameterise the
instrinsic source structure (e.g. the features on the surface of a
planet such as Titan), it is still widely applicable.

\subsection{Methods to estimate the PSF}

Below are suggested several ways one might try to infer the shape of
the PSF.
This list is not necessarily exhaustive, but is intended to
indicate that many possibilities exist if one can be a little
inventive.\\

\noindent{\em (i) Reconstruct it from the wavefront sensor data}\\
From the astronomer's perspective, this is the ideal option.
PSF reconstruction has been developed for both curvature
\cite{davies:ver97} and Shack-hartmann AO systems
\cite{davies:wei03,davies:jol04}, and there are no technical
limitations.
However, there is no
general facility for PSF reconstruction yet available at the VLT.
A tool is being developed for NACO (see Clenet's contribution to this
proceedings).\\

\begin{figure}[t]
\centering
\includegraphics[width=3cm]{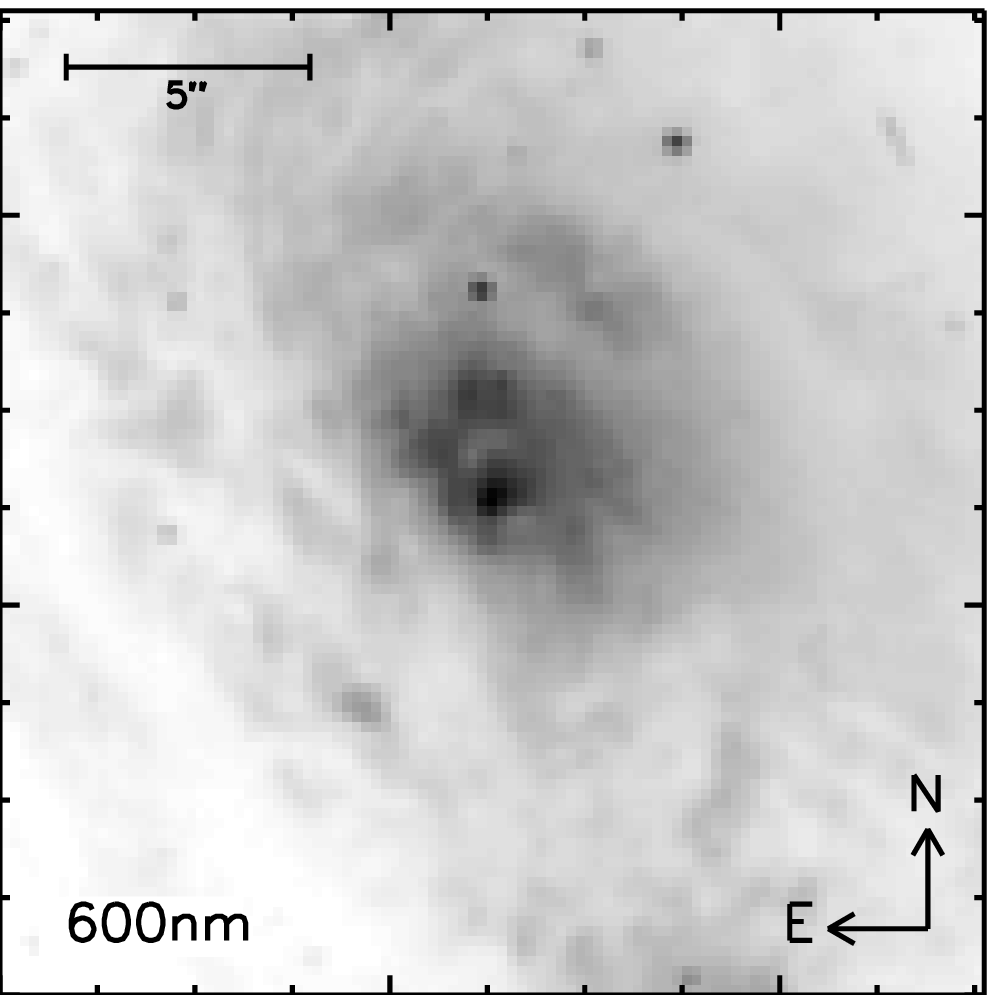}
\hspace{5mm}
\includegraphics[width=3cm]{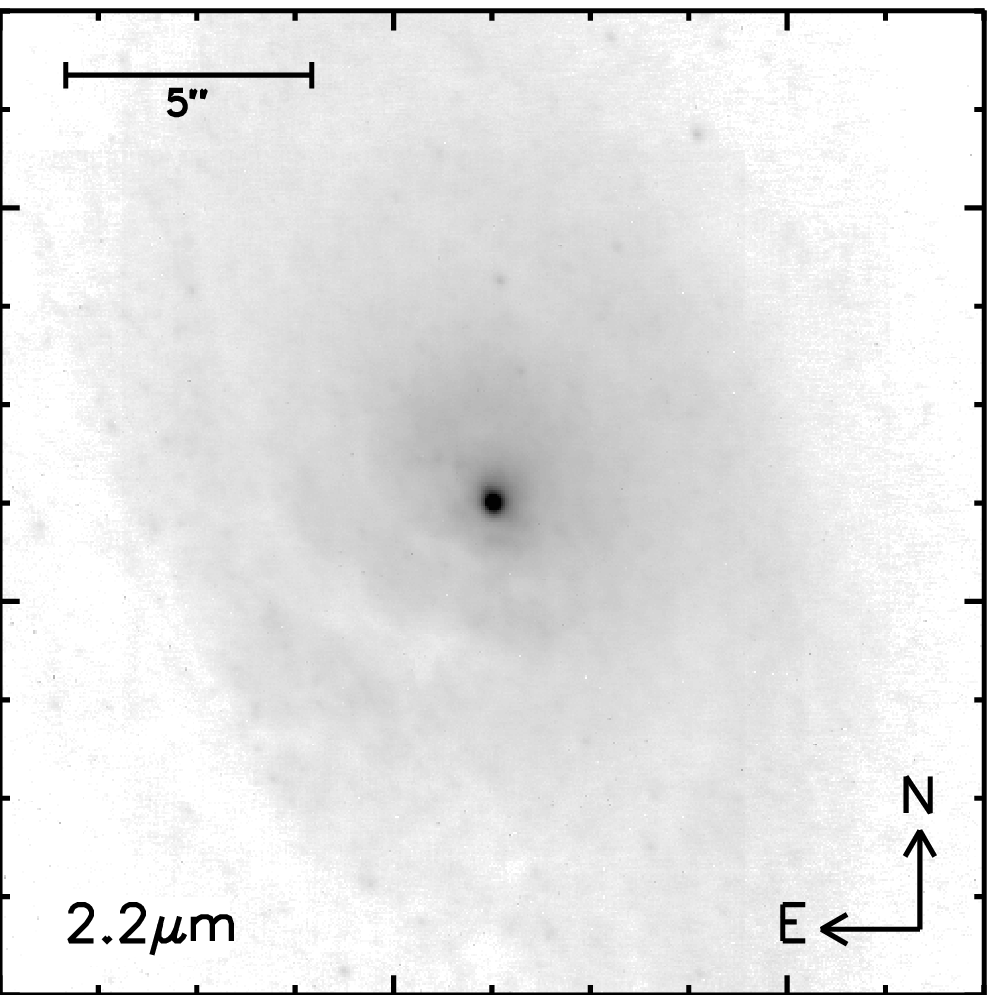}
\caption{Images of the Circinus galaxy: left optical, right near
  infrared \cite{davies:pri04}. 
It is impossible to reproduce the flux distribution seen by a visible
  WFS using a reference star.
On the other hand, what an infrared WFS sees is dominated by a point
  source and so it may be possibleto use a reference star to estimate
  the PSF.}
\label{davies:fig:circ}       % Give a unique label
\end{figure}

\noindent{\em (ii) Use an isolated star as a reference}\\
This is generally the path recommended to an observer.
However, for practical reasons -- specifically the time needed to
slew to and observe a separate reference star -- it is often impractical.
Furthermore, a reliable PSF estimate requires that the intensity and
distribution of flux on the WFS is the same for the reference star as
for the science wavefront reference object.
If both are stars, this can work well;
but as Fig~\ref{davies:fig:circ} shows for AGN and other extended
sources, it is simply unreliable.

To complicate the matter further, if the wavefront reference is not
the science target, then using it to 
estimate the PSF is misleading due to anisoplanaticism.
Instead one needs to find a pair of stars (e.g. from the Washington
Double Star Catalog \cite{davies:mas01}) separated by the same
distance, one of which matches the guide star magnitude and the other
of which can be observed by the science camera.\\

\noindent{\em (iii) Extrapolate it from surrounding stars}\\
If one is lucky, it may be possible to measure the PSF from
nearby stars \cite{davies:pri04}.
More often, it will be necessary to account for anisoplanaticism.
Several methods have been developed to estimate an off-axis PSF; 
and in principle these could be turned around to derive an on-axis PSF
from off-axis stars.
Typically, they require knowledge of the $C_N^2$ distribution
through the atmosphere \cite{davies:fus00,davies:bri06} or
observations of calibration frames containing many stars
\cite{davies:ste02}.
But it is also possible to make a reasonable (and sufficient)
approximation to the way the PSF varies across a wider field using the
science data alone, as long as at least one or two stars or compact
objects are detected \cite{davies:cre05,davies:cre06}.\\

\noindent{\em (iv) Extract it from the science data itself}\\
The broad line region in AGN is only a few lightdays across and
is therefore always unresolved in 8-m class telescopes.
In addition, the near-infrared non-stellar continuum associated with
AGN is only 1--2\,pc across and hence unresolved in AGN that are
at least $\sim20$\,Mpc away.
The spatial distribution of both these quantities can be extracted
using the spectral information available in a near infrared datacube,
and has been used as an estimate of the PSF in several cases
\cite{davies:dav04a,davies:dav04b,davies:dav06}.

One might expect that it should also be possible to extract
information about the PSF from other science data in an
analogous way.\\

\noindent{\em (v) Derive it by comparison to other higher resolution data}\\
If data taken with another instrument at another time exist at
the same wavelength and at higher spatial resolution, one
might derive the PSF by reference to these \cite{davies:mul06}.
This is because convolution of the PSF $P$ with the intrinsic source
$S$ yields the observed source $O = P \otimes S$.
One can define a broadening function $F$ which, when convolved with
the higher resolution observation $O_h$, reproduces the lower
resolution observation $O_l = O_h \otimes F$.
Then by definition the lower resolution PSF is $P_l = P_h \otimes F$.

\subsection{Effects of LGS adaptive optics}

The VLT Laser Guide Star Facility has recently been commissioned, and
so the observer will soon have to cope with LGS-AO data.
Because the wavefront reference will be the same regardless of the
science target, the PSF should in principle be easier to measure using
an isolated reference stars as in point (ii) above.
In addition, because the LGS samples a cone rather than a full column
through the atmosphere, the isoplanatic effects will be smaller.
The main impact on the PSF will be residual jitter from the tip-tilt
star, which depends on how faint and how far away it is.
But this is relatively easy to add in afterwards to an initial (better)
estimate of the PSF.

\section{Improving the Background Subtraction}
\label{davies:sec:bkg}

\begin{figure}[t]
\centering
\includegraphics[width=11.7cm]{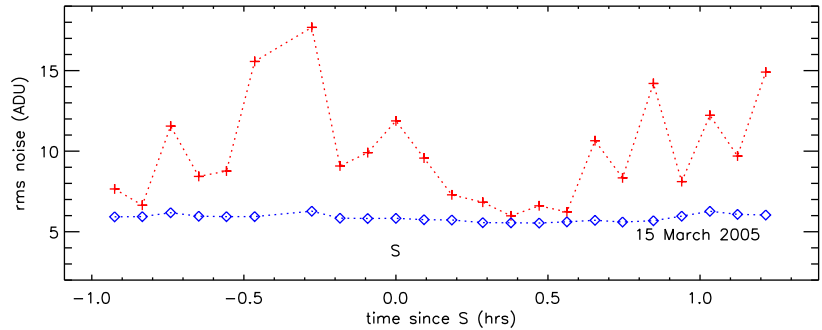}
\caption{Noise in sky subtracted H-band SINFONI cubes (arbitrary data units)
  for consecutive 5\,min integrations.
Pluses denote cubes which each had a different sky cube, taken
  immediately afterwards, subtracted.
Diamonds show the resulting noise level when this was performed using
  the scaling algorithm described in the text.}
\label{davies:fig:oh}       % Give a unique label
\end{figure}

A method to improve subtraction of the near infrared background, which
is dominated by OH emission lines, has recently been
described \cite{davies:dav07}.
This is based on the fact that most of the variation between these
lines occurs from changes in the vibrational rather than rotational
temperature of the OH radical.
And grouping the emission lines according to the
vibrational part of their transition can be done to a reasonable
approximation by wavelength: 
any particular spectral segment contains all the strong lines for
one specific vibrational transition.
One can then apply an appropriate scaling to the sky frame for
each segment separately before subtracting it.
Since the segments span a reasonably wide wavelength range, the
method is robust against emission or absorption features being blended
with OH lines.
Treatment of the rotational part of the transitions is similar,
although trickier because they cannot be grouped so easily and also
because OH lines from different transitions are blended.
The integration of this background subtraction algorithm into the
SINFONI pipeline is described by Modigliani in this proceedings.

A quantitative indication of the improvement this method can yield is
given in Fig.~\ref{davies:fig:oh}.
In this experiment, a long sequence of blank 5\,min H-band SINFONI
frames were used. 
For each frame, the successive one was used to subtract the sky
background.
The noise in the sky-subtracted cube -- which is dominated by residual
OH emission -- was measured as the standard deviation of all
data values within the spatial field and spanning 1.55-1.75\,$\mu$m.
The figure shows that the noise is variable and high.
Using this new algorithm not only reduced the residual noise by the
resulting noise is much more stable across all the frames.
In fact, this method allows one to use fewer sky frames, thus
significantly increasing the observing efficiency.

One aspect that also needs to be addressed in relation to background
subtraction is the accuracy of the wavelength calibration.
It is a feature of SINFONI data that there may be a shift of a
fraction of a pixel along the spectral axis between frames.
If this is not corrected, then subtracting a sky frame will leave
P-Cygni residuals for the OH lines.
This can be corrected by reconstructing each cube before subtracting
the background, measuring the wavelength of each strong OH line, and
deriving the mean spectral offset.
The whole frame can then be shifted by an appropriate amount to
correct the offset.
Details of all these aspects are described elsewhere
\cite{davies:dav07}.

\section{Interpolating in 3 Dimensions}
\label{davies:sec:interp}

\begin{figure}[t]
\centering
\includegraphics[width=11.7cm]{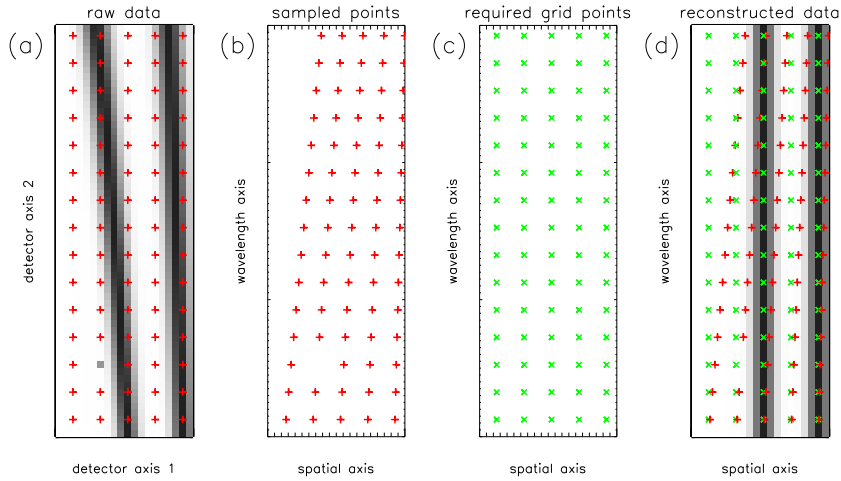}
\caption{Illustrative example of an alternative perspective for
  reconstructing datacubes.
(a) observed data are sampled regularly in the reference frame of the detector.
(b) this sampling is irregular in the reference frame of the
  reconstructed cube; bad pixels can simply be omitted from the set
  of sampled points.
(c) one can freely specify the required gridding
  (i.e. spatial/spectral pixel scale)
  for the reconstructed data; it is independent of the actual sampling.
(d) each required grid point is interpolated from sampled points which
  lie in its local neighbourhood. Any suitable algorithm can be used
  for the interpolation.
}
\label{davies:fig:interp}       % Give a unique label
\end{figure}

Interpolation is a crucial issue for integral field spectroscopy,
since reconstructing 3D datacubes requires a significant amount of
interpolation.
In the classical approach, one needs to correct for bad pixels,
straighten the spectral traces, linearise the dispersion, and finally
align the slitlets (or pixels).
Tuning the wavelength scale to correct for flexures between frames (as
described in Section~\ref{davies:sec:bkg}) can introduce an additional
interpolation step.
Poor management of the interpolation strategy or poor choice of the
interpolation scheme can degrade the quality of the final data.
For this reason, I propose an alternative perspective on the purpose
of calibrations which allows one to view the data reconstruction in a
different way -- enabling one to perform all the interpolation in a
single step while at the same time permitting a far greater
flexibility.
This can be summarised as follows:\\

\noindent{\em Standard View}\\
calibrations allow one to create the mathematical functions necessary
(e.g. polynomials) to correct the spectral and spatial curvature on
the detector.\\

\noindent{\em Alternative View}\\
calibrations allow one to create look-up tables which associate each
measured value on the detector with its spectral and spatial location
in the final reconstructed data.\\

This new perspective, which is incorporated into the design of
the KMOS data reduction library, is outlined graphically in
Fig.~\ref{davies:fig:interp}.
The most important realisation is that in `detector space' there can
be no concept of a wavelength or spatial axis.
These concepts apply only to the final reconstructed cube.
The detector is nothing more than the medium on which raw data values are
recorded.
The calibrations allow one to assign each measured value on the
detector to a spatial/spectral location in the
reconstructed cube.
Together, these locations provide an irregularly spaced sampling of
that cube.
Ihe aim is thus to reduce the raw data and the calibrations to
a list of values with their associated locations:
\[
\begin{array}{cccc}
value_0, & x_0, & y_0, & \lambda_0 \\
value_1, & x_1, & y_1, & \lambda_1 \\
\vdots & \vdots & \vdots & \vdots \\
value_n, & x_n, & y_n, & \lambda_n \\
\end{array}
\]
Data associated with bad pixels is simply excluded from the list and
so does not contribute to the set of sampled locations.
Creation of this list is the first step.
The second step is to specify the regular sampling -- i.e. the spatial
and spectral pixel size -- that is required for the reconstructed
cube.
The third step is to interpolate each of these regularly gridded
positions from sampled locations in the local neighbourhood.
In a fourth step, one can determine any spectral (or spatial) offsets
in the reconstructed cube and feed these back to create a 
new list with updated locations for each measured value.
One can then re-interpolate the regular grid of points, leading to a
final cube which has been reconstructed in a single interpolation and
which has no offsets.
There are a number of advantages of this method:
\begin{itemize}

\item
Only a single interpolation is required to reconstruct the final cube
from the raw data. This leads to improved noise properties in the
final cube.

\item
One can combine separate frames during this interpolation, 
by concatenating their lists of data values and locations. This is
useful because it avoids the need to shift and combine cubes afterwards.

\item
One has a free choice of spatial and spectral sampling in the final
cube. 
This is useful if one wants to compare the data to that from another
instrument: one can reconstruct the cube at the appropriate pixel
scale, rather than having to re-interpolate it afterwards.

\item
One has the option of smoothing the data during the reconstruction. If
the data is particularly noisy, one can increase the size of the local
neighbourhood around each point to reduce the noise at the expense of
resolution.

\end{itemize}

%%%%%%%%%%%%%%%%%%%%%%%%%%%%%%%%%%%%%%%%%%%%%%%%%%%%%%%%%%%%%%%%%%%

% BibTeX users please use
% \bibliographystyle{}
% \bibliography{}
%
% Non-BibTeX users please follow the syntax
% the syntax of "referenc.tex" for your own citations

%%%%%%%%%%%%%%%%%%%%%%%%%%%%%%%%%%%%%%%%%%%%%%%%%%%%%%%%%%%%%%%%%%%%%%  }

%%%%%%%%%%%%%%%%%%%%%%%%%%%%%%%%%%%%%%%%%%%%%%%%%%%%%%%%%%%%%%%%%%%%%%

\printindex
\end{document}